\documentclass[10pt,a4paper,twocolumn]{article}
\usepackage[a4paper,left=16mm,right=16mm,bottom=25mm,top=25mm]{geometry}
\usepackage[fleqn]{amsmath}
\usepackage{fancyhdr}
\usepackage{subfigure}
\usepackage{graphicx}
\usepackage{booktabs}
\usepackage[british]{babel}
\usepackage[square,comma,numbers,sort&compress]{natbib}
\usepackage{csvsimple}
\usepackage{graphicx}

\usepackage{tikz}
\usepackage{array}
\usepackage{tabu}
\usepackage{multirow}
\usepackage{url}
\usepackage{xcolor}
\usepackage{afterpage}
\usepackage{graphicx}% 
\usepackage{lipsum}
\usepackage{bm}% 
\usepackage[title]{appendix}
\usepackage[]{subfigure}
\usepackage{float}
\usepackage{hyperref}
\hypersetup{colorlinks,linkcolor={blue},citecolor={blue},urlcolor={cyan}}
\usepackage{cleveref}
\usepackage{afterpage}
\usepackage{caption}
\usepackage{lineno}
\usepackage{lettrine}
\usepackage{color}
\usepackage{soul}
\usepackage{ulem}
\usepackage{float}
\usepackage{placeins}

\definecolor{greendark}{rgb}{0.0,0.5,0.0}

\usepackage{hhline}
\usepackage{multirow}

\usepackage{graphicx}% Include figure files

\begin{document}
\title{\Large{Interlayer Magnetic Coupling in FePS$_{3}$ and NiPS$_{3}$ Stacked Bilayers}}

\author{Andrea  Le\'on,$^{1,4,*}$ Beatriz Costa,$^{2,3,*}$ Thomas Heine,$^{2,3,4,5}$ Thomas Brumme$^4$}

\date{%
$^1$Departamento de F\'{i}sica, Facultad de Ciencias, Universidad de Chile, Casilla 653, Santiago, Chile. \\%
$^2$Helmholtz-Zentrum Dresden-Rossendorf, Bautzner Landstr. 400, 01328 Dresden, Germany.\\
$^3$Center for Advanced Systems Understanding, CASUS, Untermarkt 20, 02826 Görlitz, Germany.\\
$^4$Chair of Theoretical Chemistry, Technische Universit\"at Dresden, Bergstrasse 66, 01069 Dresden, Germany.\\
$^5$Yonsei University and ibs-cnm, Seodaemun-gu, Seoul 120-749, Republic of Korea.\\
$^*$These authors contributed equally to this work.\\
\today
}

\twocolumn[
\begin{@twocolumnfalse}
\maketitle
\begin{abstract}
Single layers of transition-metal thiophosphates (2D-TMPS$_{3}$) van der Waals magnets are an ideal platform for studying antiferromagnetic interactions in 
two dimensions. 
However, the magnetic coupling mechanism between two or more individual layers of these materials remains mostly unexplored.
This study presents a density-functional based analysis and analytical models to describe the magnetic configurations of FePS$_{3}$ and NiPS$_{3}$ stacked bilayers. We explore the interplay between magnetic configurations and stacking shift, therefore identifying the mechanisms that result in either ferromagnetic or antiferromagnetic coupling between layers. Our findings indicate that the stacking with the lowest energy is metal-dependent, and the interlayer magnetic configuration (ferromagnetic or antiferromagnetic) varies based on the stacking type and the metal involved. Using an Ising-Hamiltonian model and a tight-binding model based on Wannier functions, we show that interlayer exchange interactions must be considered up to the third nearest neighbor and to elucidate the superexchange mechanism for the NiPS$_{3}$ system.\\
\end{abstract}

\end{@twocolumnfalse}]

\section{Introduction}
\footnotetext[1]{andrea.leon@uchile.cl}
\footnotetext[2]{thomas.brumme@tu-dresden.de}

\lettrine[lines=2, loversize=0.25, findent=2pt]{\LettrineFont{S}}tacking engineering in van der Waals (vdW) layered magnetic systems offers new ways to manipulate and study magnetic phases \cite{xie2022twist}. Techniques such as layer shifting, rotation, moir\'e patterns, and forming heterostructures by stacking different layers are currently employed in this field \cite{ghader2020magnon}. Particularly, the exploration of stacked antiferromagnetic (AFM) and semiconducting materials is revealing new physical phenomena with potentially transformative impacts \cite{rahman2021recent,NiPS3}. The 2D-TMPS$_{3}$ (TM $=$ transition metal atom) systems being semiconductors and antiferromagnets are garnering attention for their unique optical properties coupled with their magnetic ordering. Currently, utilizing stacking degrees of freedom to merge these properties opens new avenues for investigating magnetic phenomena and interface effects \cite{ramos2022photoluminescence,bora2021magnetic}. In this context, research focused on unraveling the role of AFM order, stacking degrees of freedom, and the role of the TM atoms is necessary for further understanding and controlling desired properties in the search for functional materials.

The first-row TMPS$_{3}$ (TM = Mn, Fe, Ni, and Co) family exhibits diverse AFM properties \cite{MPS3_review}. They can behave like Heisenberg (MnPS$_{3}$ \cite{MnPS3}), Ising (FePS$_{3}$ \cite{FePS3}), and XY-type (CoPS$_{3}$ \cite{CoPS3} and NiPS$_{3}$ \cite{NiPS3}) antiferromagnets. Besides their magnetic properties, these materials show intriguing optical properties that vary from bulk to layered structures, including optical excitations ranging from near IR to UV \cite{ramos2021ultra} and excitons with large binding energies \cite{wyzula2022high,mai2021magnon}. They can be exfoliated into a few atomic layers, enabling studies on magnetic ordering and dimensionality \cite{thickness_MnPS3,thickness_FePS3}. However, detailed measurements are challenging due to the complexity of experimental setups for materials with zero total magnetization \cite{kim2019suppression}. These materials' magnetic, electronic, and optical properties offer an exciting landscape for optoelectronic and spintronic applications \cite{yan2021correlations,ramos2022photoluminescence,duan2021enhanced}.

\begin{figure}
\includegraphics[width=0.49\textwidth]{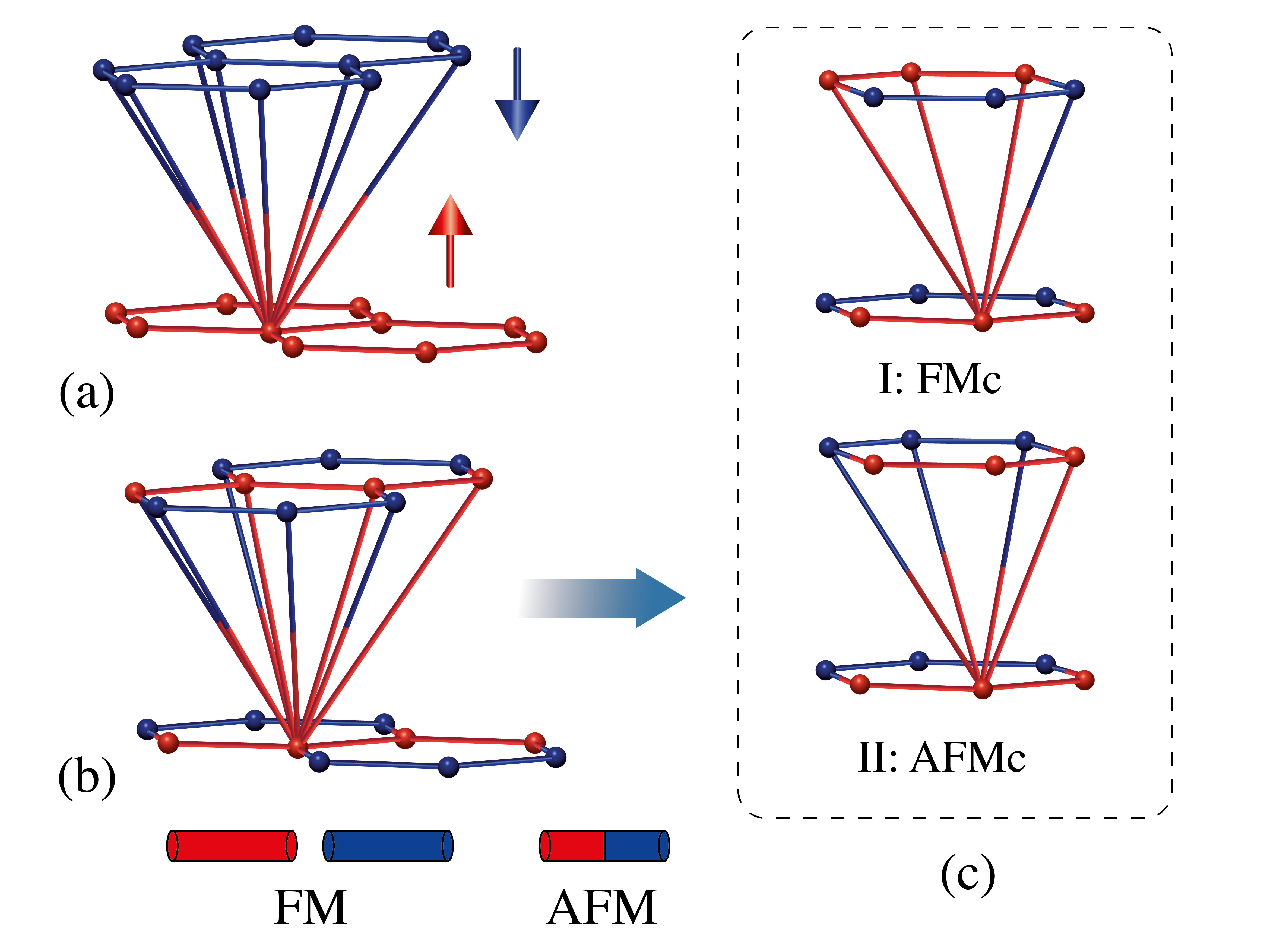}
\caption{Illustration of two possible configurations for achieving AFM bilayers: (a) by coupling two FM layers of opposite spin, and (b) stacking two AFM layers. (c) Configuration I (I) and Configuration II (II), which denotes two possible magnetic coupling between the layers, defined as: ferromagnetic and antiferromagnetic interaction between first neighbors.}
\label{00fig0}
\end{figure}

TMPS$_{3}$ monolayers exhibit magnetic properties dictated by superexchange interactions between metal nearest neighbors (NN)\cite{chittari2016electronic,autieri2022limited}. While intralayer interactions are well-studied, interlayer interactions remain relatively unexplored \cite{kim2021magnetic}. In other bilayer systems, e.g, the CrX$_{3}$ family \cite{sivadas2018stacking,jang2019microscopic}, the magnetic interlayer interactions are influenced by the contributions of first, second, or even third neighbors, requiring mechanisms beyond superexchange to be understood. Unlike FM layers, the stacked AFM sheets cannot be simply defined with FM or AFM coupling (denoted as FMc and AFMc, see Fig.~\ref{00fig0}). Depending on the in-plane magnetic configuration (N\'eel, zigzag, or stripy), the NN interactions between layers can be AFMc or FMc, which could lead to magnetic competition or spin frustration. This frustration depends on the number of metal NN, distances, and the superexchange mechanism involved among the metal and sulfide atoms of the bottom and top layers. Models that consider distance and hopping processes are necessary to understand the mechanism behind of the magnetic interactions in AFM bilayers.\\

While interlayer interactions in TMPS$_{3}$ systems are generally weaker than intralayer interactions, significant interlayer effects vary with the TM type \cite{chu2020linear}. For example, strong interlayer interactions in NiPS$_{3}$ are evident in processes like layer exfoliation and under pressure \cite{ma2021dimensional}. Conversely, Raman studies suggest weaker interlayer couplings in FePS$_{3}$ and MnPS$_{3}$, with FePS$_{3}$ showing stronger effects than MnPS$_{3}$ \cite{yan2022layer}. However, systematic theoretical studies on the influence of layering are lacking, which are crucial for manipulating AFM magnetic phases.\\

In this work, we study the influence of stacking on the magnetic coupling between layers and the electronic properties in bilayer FePS$_{3}$ and NiPS$_{3}$. First, we investigate the total energy of the bilayer systems to establish the magnetic ground state. Second, we computed the exchange constant using the classical Ising model. For a second method based on Green's functions, we constructed Wannier functions from first principles. Then, we applied a spin rotation to Green's functions, providing a tight-binding (TB) Hamilçonian model to calculate the exchange constant. Our findings indicate that the most stable stacking and the magnetic coupling between layers depend on the metal atom. Through the solution of the Ising and TB Hamiltonian in the bilayers, we find that the most substantial interlayer exchange interaction occurs for the second or third NN, depending on the stacking and the metal atom. We focus on NiPS${3}$, which exhibits stronger exchange interactions than FePS${3}$, in agreement with experimental findings \cite{ma2021dimensional,yan2022layer}. With the advent of Wannier orbitals, we find that the contributions to the interlayer magnetic exchange are determined by the alignment or not of the $p$ orbitals of the sulfur atoms in each layer.

\section{Computational details}

We used density functional theory (DFT) with the Perdew-Burke-Ernzerhof (PBE) generalized gradient approximation functional \cite{PBE}, implemented in the Vienna ab initio Simulation Package (VASP) \cite{VASP}, for our theoretical analysis using the plane-wave pseudopotential method. To account for London dispersion forces, we have employed the van der Waals (vdW) approximation optB86b-vdW \cite{optB86b}. Using this approach, we have achieved a good agreement with the lattice parameters and interlayer distances as reported in the bulk \cite{jenjeti2018field,yan2022layer,pazek2024charge} (see, supplemental material (SM) \ref{SM-bulk}).
A Hubbard on-site Coulomb parameter of U = 5 eV is used for the Fe and Ni atoms to accurately take into account the electronic correlations \cite{haddadi2024site} along the Dudarev approach \cite{dudarev1998electron}. Additionally, we have performed calculations with U = 3.0, 4.0, 6.0 
eV to study the influence of the Hubbard-U repulsion on specific properties. For the structural optimization of the bilayers, we include 25 \AA{} of vacuum space to minimize the interaction between periodically repeated images along the z-axis. The full structural optimizations including the lattice of the bilayer systems are performed with a force convergence of 10$^{-3}$ eV/\AA$^{-1}$ for each atom and a plane-wave energy cutoff of 500 eV. Regular Monkhorst-Pack \cite{Monkorst} grids of 11$\times$5$\times$1 for the atomic relaxation, and  15$\times$9$\times$1 for the self-consistent calculation have been used.

The exchange parameters were calculated through the Ising model and the TB2J python package \cite{tb2j}. For the second method, a tight-binding model using localized Wannier functions (WF) as implemented in the Wannier90 code \cite{Wannier90} is derived from the DFT results. More details about the Wannierization can be found in \ref{Wannier} section.  
\begin{figure}
\includegraphics[width=0.45\textwidth]{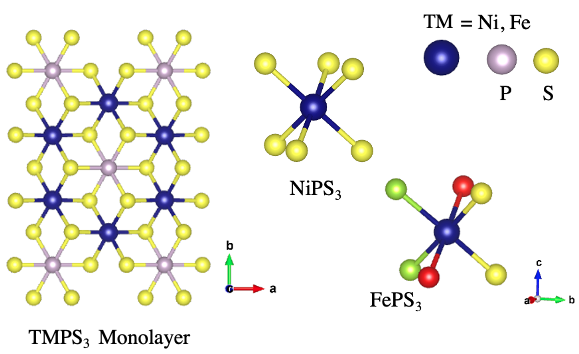}
\caption{Monolayer structure for TMPS$_{3}$, TM= Fe and Ni. NiPS$_{3}$ exhibits a nearly perfect octahedral environment, while FePS$_{3}$ displays a distorted octahedra due to the different Fe-S bonding. The Fe-S color bonds symbolize the non-equivalent length. }
\label{fig00}
\end{figure}

\section{Results}

\begin{figure*}[ht]
\centering
\includegraphics[width=0.9\textwidth]{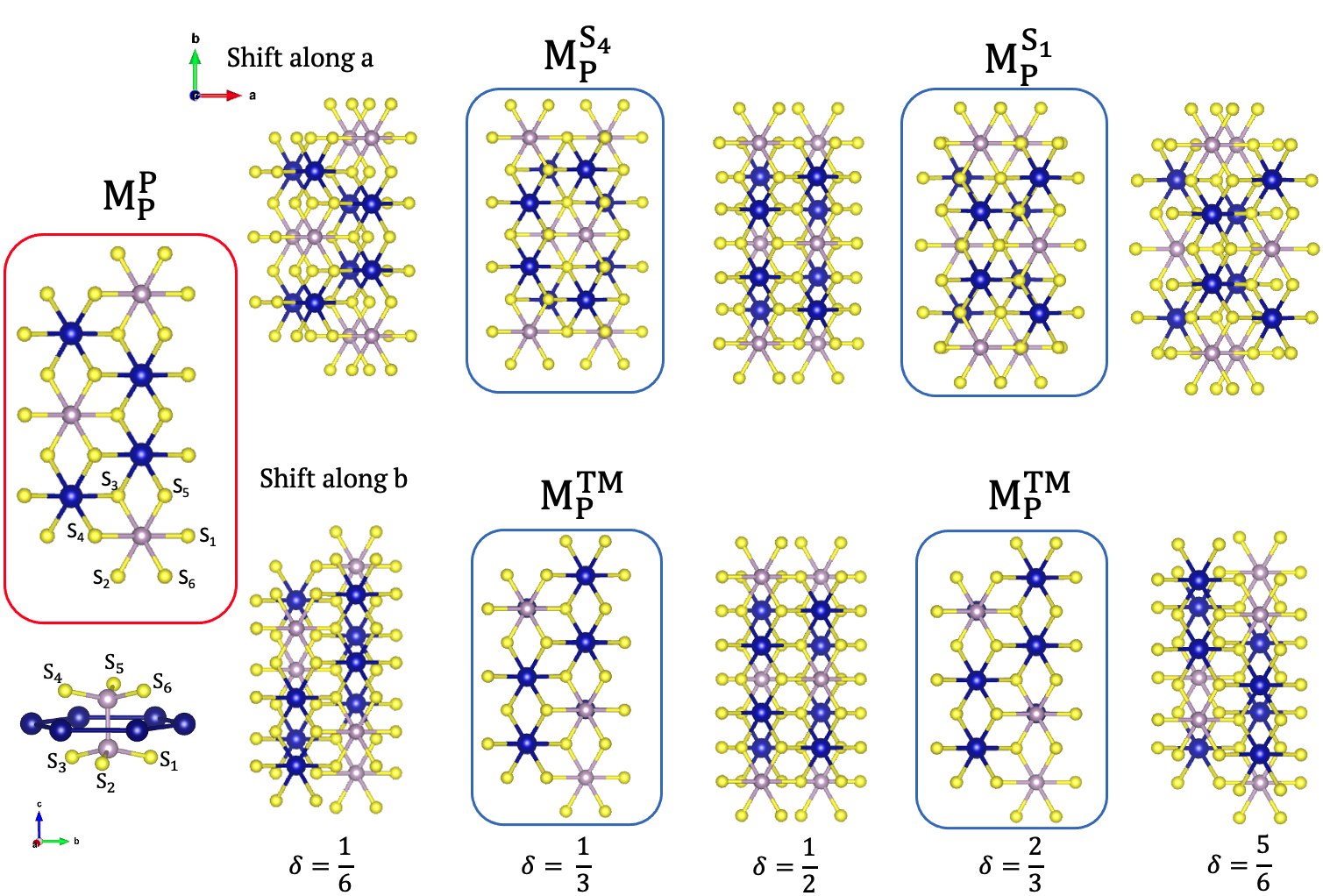}
\caption{Bilayer structures for the TMPS$_{3}$ systems under shift along $\vec{a}$ (upper panel) and $\vec{b}$ directions (lower panel). The relative shift correspond to  $\delta$ = $\frac{1}{6}$, $\frac{1}{3}$, $\frac{1}{2}$, $\frac{2}{3}$ and $\frac{5}{6}$. The systems in the square box correspond to the high symmetry stackings.}
\label{fig0}
\end{figure*}

Bulk TMPS$_{3}$ crystallizes in a monoclinic structure with centrosymmetric C2/m point symmetry. The monolayers are then stacked in the c direction in an AB manner. Since two P atoms and six S atoms are covalently bonded among themselves, forming a (P$_{2}$S$_{6}$)$^{4-}$ anion complex, each transition metal has a formal charge of $+2$. NiPS$_{3}$ and FePS$_{3}$ exhibit inversion symmetric zigzag AFM order~\cite{kim2019suppression, hwangbo2021highly,wildes2015magnetic,belvin2021exciton,lee2016ising,lanccon2016magnetic,geraffy2022crystal}. Both systems have trigonally distorted MS$_{6}$ octahedra forming an edge-sharing layered honeycomb lattice in the $\mathbf{a}/\mathbf{b}$ plane. In contrast to NiPS$_{3}$, FePS$_{3}$ displays enhanced trigonally distorted FeS$_{6}$ octahedra (Fe-S bonds are not equivalents, see Fig.~\ref{fig00}) \cite{murayama2016crystallographic}\cite{afanasiev2021controlling}. 

First, we analyzed several stackings of bilayer TMPS$_{3}$. These stackings were constructed starting with the fully relaxed monolayers, where the upper layer is rigidly shifted by a shift $\delta$ (in units of the lattice constant) along $\mathbf{a}$ ([100]) and $\mathbf{b}$ ([010]). Fig.~\ref{fig0} shows the non-equivalent high-symmetry stackings corresponding to shifts in which the sulfur atoms are above the metal or around the P dumbbell. The nomenclature of the stackings is similar to the one introduced by Yu \textit{et al.} for transition-metal dichalcogenides \cite{yu2018}. The different stackings are labeled M$^{a}_{b}$, where M stands for bulk-like monoclinic stacking without rotation between layers, and $a$ and $b$ indicate the high-symmetry positions in the upper and lower layers, respectively. Shifts along [010] direction can result in non-monoclinic structures; however, we continue to use M for simplicity.

\begin{figure*}[ht]
\centering
\includegraphics[width=0.4\textwidth]{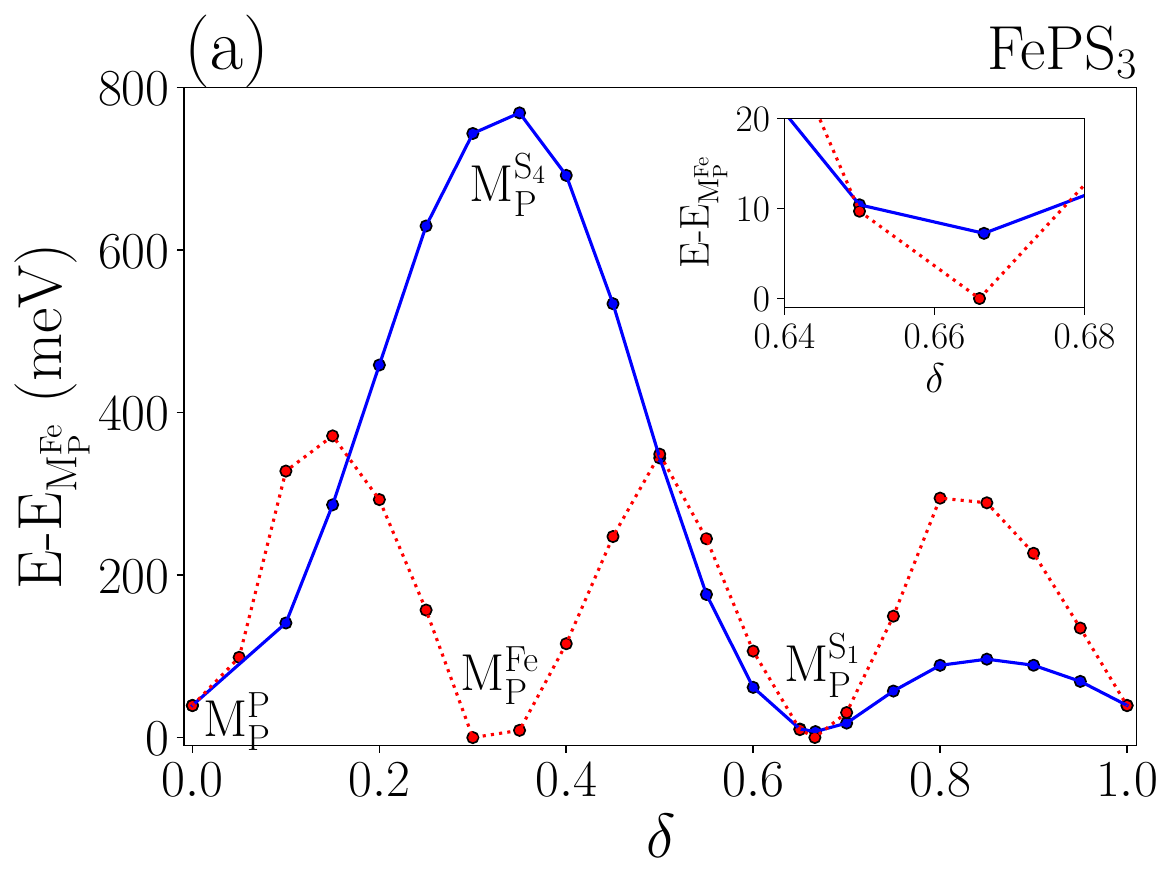}\includegraphics[width=0.4\textwidth]{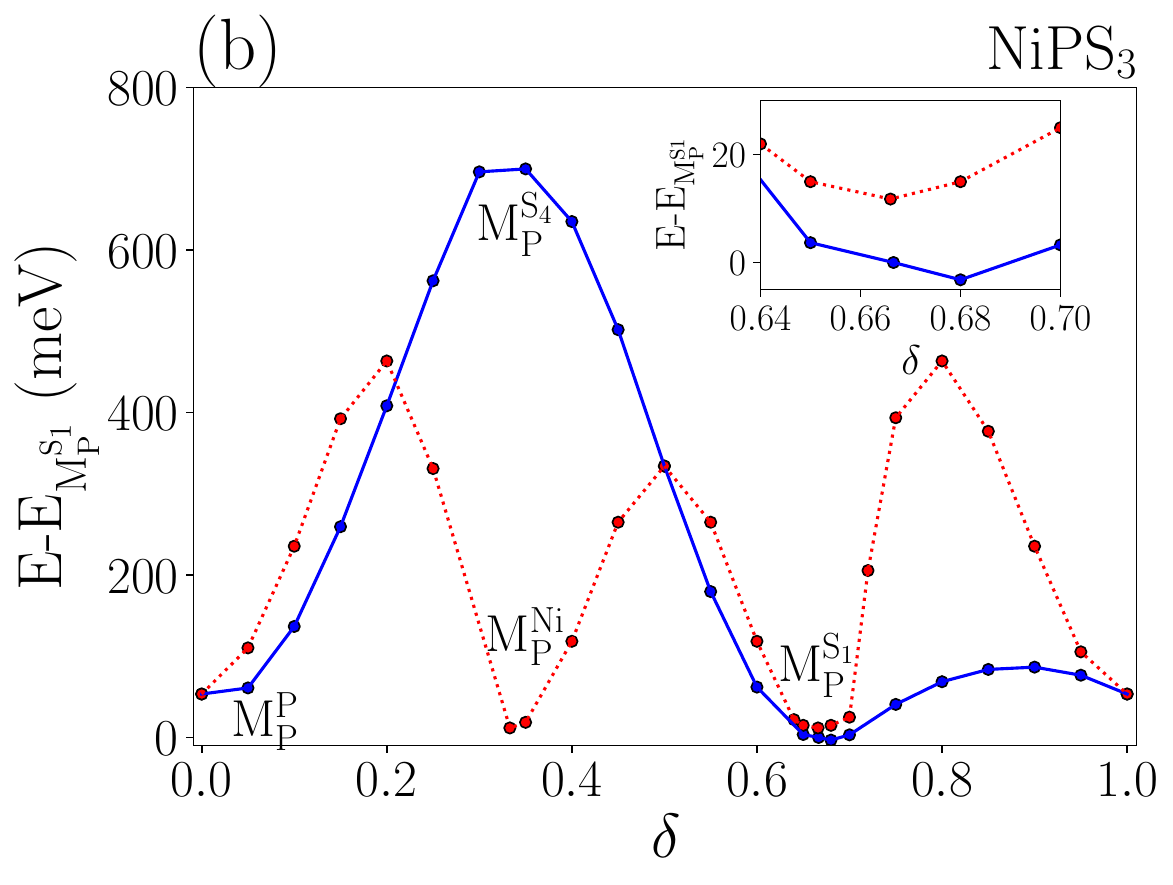}
\includegraphics[width=0.4\textwidth]{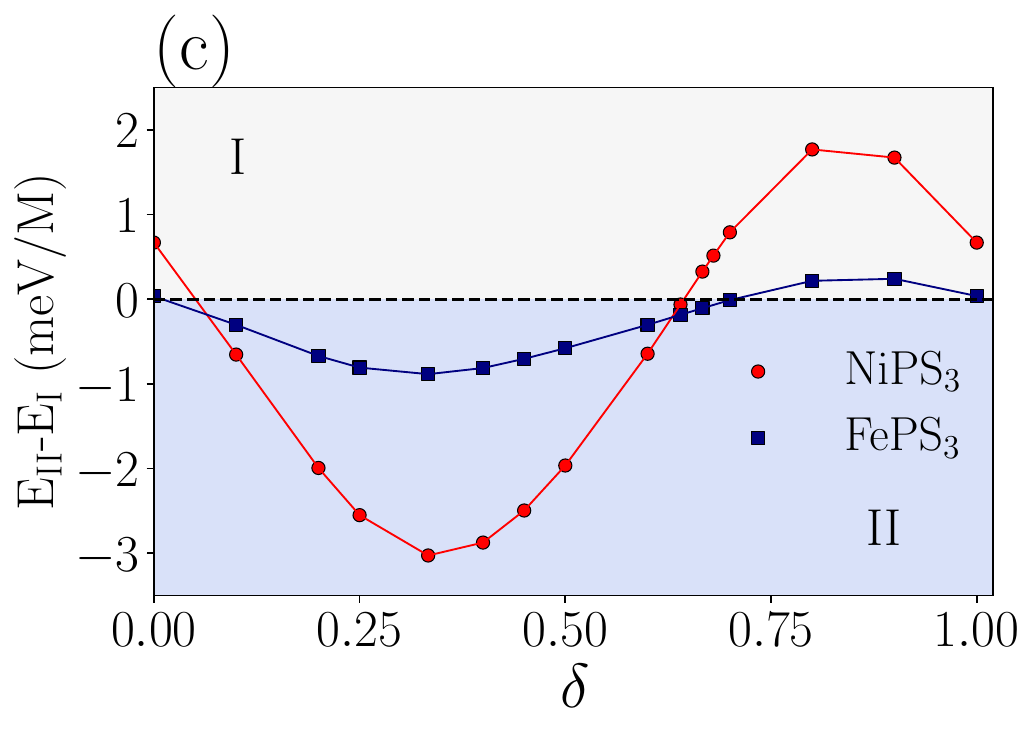}
\raisebox{0.5cm}{\includegraphics[width=0.40\textwidth]{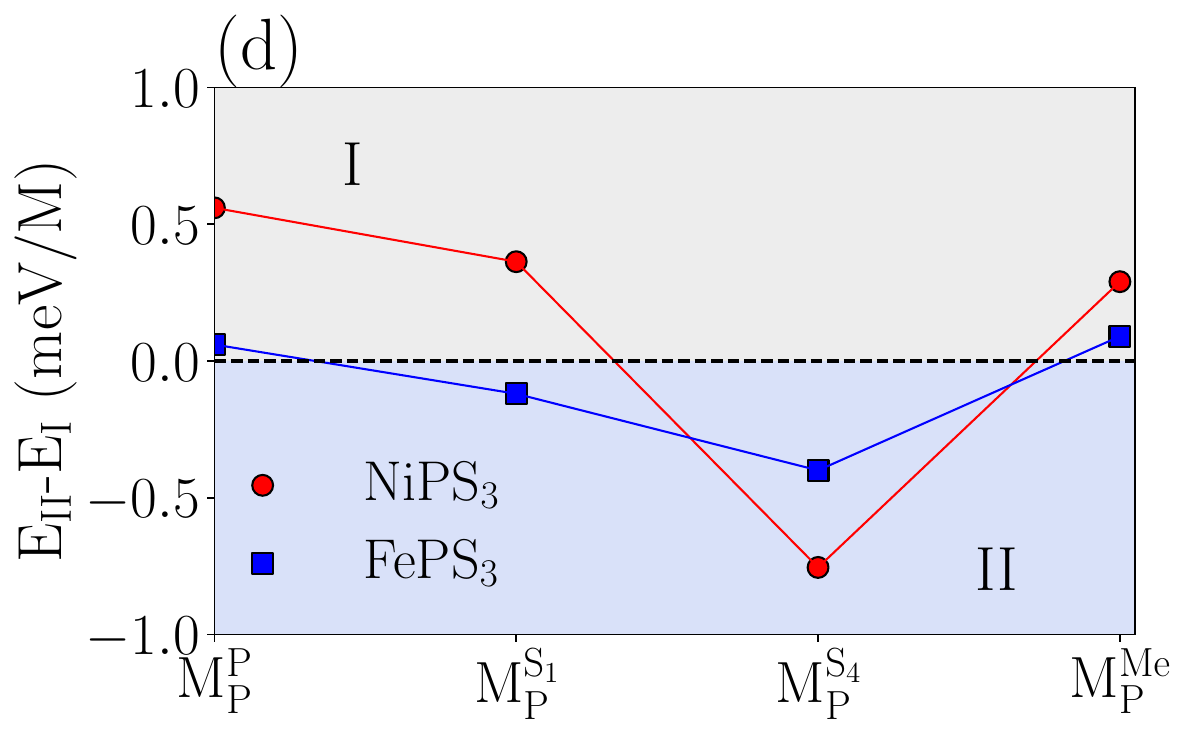}}
\begin{tikzpicture}
    \node at (1, 0.5) {\Large [100] };
    \draw[blue, ultra thick] (0,0) -- (2,0);    
    \node at (4, 0.5) {\Large [010]};
    \draw[red, ultra thick, dashed] (3,0) -- (5,0);
\end{tikzpicture}
\caption{(a)-(b) Stacking energy as a function of lateral shift ($\delta$) with respect to the ground state stacking M$_{P}^{TM}$ and M$_{P}^{S_{1}}$ for FePS$_{3}$ and NiPS$_{3}$, respectively. The red and blue lines indicate the displacement along the [100] and [010] directions, respectively.
(c) and (d) Total energy difference between configuration II and I (see Fig.\ref{00fig0}c) as a function $\delta$ along [100] and for selected high symmetry stacking, respectively (optimized structures). FMc and AFMc mean regions with AFM and FM coupling, respectively. 
}
\label{fig2}
\end{figure*}

Figures~\ref{fig2} (a) and (b) show the energy difference among different stackings relative to the ground state energy for FePS$_{3}$ and NiPS$_{3}$, respectively. Both systems exhibit similar behavior, with comparable local minima and global and local maxima. For NiPS$_{3}$, M$_{P}^{S_1}$ is the ground state, which is the bulk-type stacking; however, for FePS$_{3}$, the M$_{P}^{Fe}$ stacking (P-3 point symmetry) is favored, being lower in energy by 9.42 meV/formula unit (f.u.) (see Fig.~\ref{fig2}(a), insert figure). The dependence of the ground state on the metal atoms is also observed in other TMPS$_{3}$; for example, MnPS$_{3}$, which presents M$_{P}^{P}$ as the lower energy stacking \cite{li2013coupling}. Since we used the Hubbard U parameter to describe electron correlations in the 3d orbitals, we verified that different U values (see SM  \ref{Uvalues} for Hubbard-U = 4, 5, and 6 eV) yield consistent results regarding the difference in stackings. For the  FePS$_{3}$ system the binding energies, computed by the difference between the total energies of the optimized bilayer and monolayer, E$_{BL}$-2E$_{ML}$, are -26.35 meV/f.u and -26.60 meV/f.u for M$^{S_{1}}_{P}$  and M$^{Fe}_{P}$ stacking, respectively, which is close to the average thermal energy at 300 K, indicating that both stackings could be feasible at room temperature.\\ 

The shift of the layers relative to each other leads to a change in the interlayer distance (in the presence of lattice relaxation), as shown in Table \ref{table1} (where distance is defined as the difference in the z coordinates of the S atoms facing each other, see \ref{front-bilayer}, SM). The main effect of the relaxation in the different stackings is a slight change in the distances between layers, which is typical for most 2D materials \cite{li2024relaxation}. However, an exception occurs in the M$^{S_{4}}_{P}$ stacking, where the layers are moved farther apart. This structural change is due to the strong S-S interlayer interaction, as the S atoms are positioned directly above one another (see \ref{front-bilayer}, SM). This arrangement results in a strong Coulomb repulsion that surpasses the van der Waals interactions.

Furthermore, it is notable that NiPS$_{3}$ exhibits a slightly larger interlayer distance compared to the FePS$_{3}$ system. To discern the effects of structural distortions present in FePS$_{3}$ and the intralayer distance on the stability of the M$^{Fe}_{P}$ stacking over M$_{P}^{S_{1}}$ in this system, we conducted frozen cell calculations for FePS$_{3}$ using the relaxed structure of NiPS$_{3}$. This resulted in FePS$_{3}$ having a nearly isotropic structure, similar to the comparison made between FePS$_{3}$ and MnPS$_{3}$ in Ref.~\cite{geraffy2022crystal}. In our case, the M$^{Fe}_{P}$ stacking still exhibits lower energy by 11.18 and 8.75 meV per cell, considering the interlayer distances of $d = 2.33$ Å and 2.40 Å, respectively. Hence, the stability of M$_{P}^{Fe}$ remains independent of layer distances and M-S bond distortion.

Bulk FePS$_{3}$ and NiPS$_{3}$ exhibit AFM zigzag order within their layers and ferromagnetic or antiferromagnetic coupling between the layers \cite{chu2020linear}. 
We conducted calculations for different spin arrangements within each stacking to explore the stacking effect on the magnetic configuration between layers. For this purpose, we defined Configuration I (I) and Configuration II (II), with ferromagnetic and antiferromagnetic coupling (FMc and AFMc, respectively) between the first neighbors (see Fig.~\ref{00fig0}c, for bulk type, M$_{P}^{S_{1}}$, stacking).\\

Fig.~\ref{fig2}(d) shows the magnetic exchange energy, defined as the energy difference between the system in II and I (E$_{II}$ - E$_{I}$)\cite{sivadas2018stacking}. As $\delta$ increases along the [100] direction, a interplay between stacking and magnetic coupling is evident in both systems. In general, a local monotonic behavior is observed, where for small and large shifts, both systems favor AFMc and FMc between layers, respectively. One point of interest is the bulk-type stacking (M$_{P}^{S_1}$) located at $\delta = \frac{2}{3}$. In this transition zone, FePS$_{3}$ favors AFMc, with an energy difference $\Delta$E of -0.15 meV/Fe, while NiPS$_{3}$ favors FMc, with a $\Delta$E of 0.25 meV/Ni.\\

The ionic relaxation slightly modifies the magnetic exchange energy for M$_{P}^{S_1}$ and M$_{P}^{Fe}$ (see Fig.~\ref{fig2}(d)). However, for M$_{P}^{S4}$ stacking, the relaxed structures have the layers farther apart (see Table~\ref{table1}), resulting in less Coulomb repulsion energy and decreasing exchange energy. For $\delta$ along the [010] direction, a similar behavior is observed (see the discussion in the SM, Section~\ref{magconf}). Additionally, we have verified the robustness of the interlayer configuration by changing the distance between the layers (in the range  d$\pm$0.25\AA ).\\

\begin{table}[h]
\begin{center}
\resizebox{8.4cm}{!}{%
\begin{tabular}{|c|cc|cc|}
\hline
& \multicolumn{2}{c|}{FePS$_{3}$}                          & \multicolumn{2}{c|}{NiPS$_{3}$}                          \\ \cline{2-5} 
\begin{tabular}[c]{@{}c@{}}Stacking\\ (St)\end{tabular} & \multicolumn{1}{c|}{$\Delta$E$^{St}$ (meV/f.u.)} & \begin{tabular}[c]{@{}c@{}}Distance (\AA)\\ S$^{top}$-S$^{down}$\end{tabular} & \multicolumn{1}{c|}{ $\Delta$E$^{St}$ (meV/f.u.)} & \begin{tabular}[c]{@{}c@{}}Distance (\AA)\\ S$^{top}$-S$^{down}$\end{tabular} \\ \hline
M$^{S1}_{P}$                                                    & \multicolumn{1}{c|}{6.98}       & 2.33                                                        & \multicolumn{1}{c|}{0}       & 2.37                                                        \\
M$^{TM}_{P}$                                                        & \multicolumn{1}{c|}{0}          & 2.30                                                         & \multicolumn{1}{c|}{12.63}          & 2.36                                                         \\
M$^{P}_{P}$                                                       & \multicolumn{1}{c|}{37.8}       & 2.328                                                        & \multicolumn{1}{c|}{58.79}       & 2.40                                                       \\
 M$^{S4}_{P}$                                                       & \multicolumn{1}{c|}{344.57}     & 2.820                                                        & \multicolumn{1}{c|}{335.36}     & 2.81                                                      \\ \hline
\end{tabular}%
}
\caption{Stacking energy, $\Delta$E$^{St}$, for FePS$_{3}$ and NiPS$_{3}$ given by $\Delta$E$^{St}$ = E$^{St}$-M$_{P}^{TM}$ and  E$^{St}$ - M$_{P}^{S_{1}}$, respectively (optimized structures).} 
\label{table1}
\end{center}
\end{table}
\begin{figure}[ht]
\includegraphics[width=0.245\textwidth]{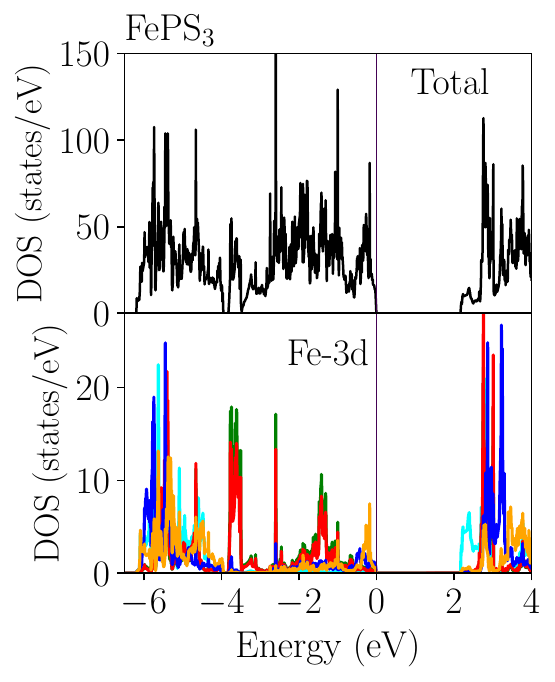}\includegraphics[width=0.225\textwidth]{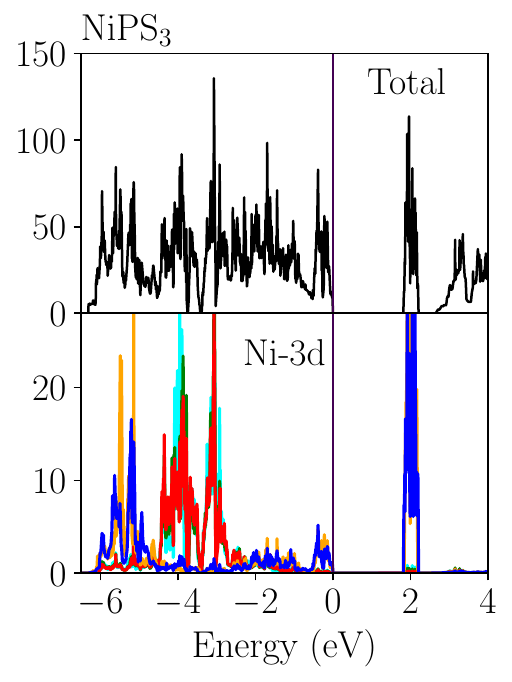}
\centering
 \begin{tikzpicture}
    \node at (0.5, 0.5) {\large $d_{xz}$};
    \draw[greendark, ultra thick] (0,0) -- (1,0);
    
    \node at (2, 0.5) {\large $d_{xy}$};
    \draw[cyan, ultra thick] (1.5,0) -- (2.5,0);
    
    \node at (3.5, 0.5) {\large $d_{yz}$};
    \draw[red, ultra thick] (3,0) -- (4,0);
    
    \node at (5, 0.5) {\large $d_{z^2}$};
    \draw[orange, ultra thick] (4.5,0) -- (5.5,0);
    
    \node at (6.5, 0.5) {\large $d_{x^2-y^2}$};
    \draw[blue, ultra thick] (6,0) -- (7,0);
\end{tikzpicture}
\caption{Density of states (DOS), total and projected to TM atoms for M$^{S_{1}}_{P}$ (bulk type stacking).}
\label{fig:dos}
\end{figure}

The stacking dependence observed in FePS$_{3}$ and NiPS$_{3}$ offers an effective platform for studying the mechanism behind the magnetic coupling in TMPS$_{3}$ systems. FePS$_{3}$ and NiPS$_{3}$ exhibit similar behavior, with the main difference being that NiPS$_{3}$ displays a stronger magnetic exchange energy, indicating that it may host strong interlayer interactions, in agreement with experimental reports \cite{ma2021dimensional}. These differences arise from the different occupation of the 3$d$ states, leading to distinct superexchange mechanisms. Figure~\ref{fig:dos} shows the density of states for each system. In NiPS$_{3}$, a gap opens between the e$_{g}$ and t$_{2g}$ states (fully occupied and almost degenerate). In FePS$_{3}$, the t$_{2g}$ orbitals split, lowering the energy of the d$_{xy}$ states while maintaining the degeneracy of the d$_{xz}$ and d$_{yz}$ orbitals. Additionally, the e$_{g}$ states split. These results are consistent with previous experimental studies proposing that the t$_{2g}$ states split into a$_{1g}$ and e${g_{\pm}^{\pi}}$ states, defined as linear combinations of d$_{xy}$, d$_{yz}$, and d$_{xz}$ states \cite{chang2022trigonal}. However, this disagrees with theoretical calculations finding orbital ordering due to the occupation of d$_{x^{2}-y^{2}}$ orbitals \cite{amirabbasi2023orbital}. These differences may arise from the definition of the octahedra-axis reference, since the orbital projection in Ref.~\cite{amirabbasi2023orbital} is not made with octahedra aligned to the Cartesian coordinate system. In the next section, we will thus focus on understanding the exchange mechanism between the layers, mainly in the NiPS$_{3}$ system -- more details for FePS$_3$ can be found in the SI.

\subsection{Magnetic exchange}

Our interest lies in stacking-dependent magnetism and understanding the mechanism behind the interplay between magnetic coupling and stacking.
Therefore, we will focus on the interlayer exchange interaction (J$^{inter}$), which can be determined by solving the effective classical Hamiltonian on a honeycomb lattice:
\begin{equation}
H = E_{0} -\frac{1}{2} \sum_{i,j} J^{intra}_{ij} \vec{S_{i}} \cdot \vec{S_{j}} - \frac{1}{2} \sum_{i,j} J^{inter}_{ij} \vec{S_{i}} \cdot \vec{S_{j}}
\label{ising1}
\end{equation}
where S$_{i}$ is the total spin at atomic site $i$, J$_{ij}$ are the exchange coupling parameters between two local spins, and $E_0$ is the total energy of the nonmagnetic state. We solve Eq.~(\ref{ising1}) for five magnetic configurations, considering the first three intralayer and interlayer neighbors for the lower-energy stackings (M$_{P}^{S_{1}}$ and M$_{P}^{Ni}$). The solutions for M$_{P}^{S_1}$ stacking is given by:
\begin{align}
E_\mathrm{N\acute{e}el}- E_{0} &= \frac{NS^{2}}{2}(3J_{1}-6J_{2}+3J_{3}) \notag,\\
E_{z}^{I/II} -  E_{0} &=  \frac{NS^{2}}{2}((-J_{1}+2J_{2}+3J_{3}) 
                     \pm (-2J_{4}+2J_{6})),\notag\\
E_{z/\mathrm{ferri}} - E_{0}&=\frac{NS^{2}}{4}(-J_{1}+3J_{3}-2J_{4}+2J_{6})\notag,\\
E_{s/\mathrm{ferri}}^{I/II}- E_{0} &= \frac{NS^{2}}{4}(J_{1}-3J_{3}
           \pm (-2J_{5}+2J_{6}))\notag,\\
E_\mathrm{ferri}^{I}- E_{0} &= \frac{NS^{2}}{2}(-2J_{2}+J_{4}+J_{5}),\notag\\
\label{eq1}
\end{align}
and for the M$_{P}^{Ni}$ stacking by:
\begin{align}
E_\mathrm{N\acute{e}el}^{I/II} -E_{0} &= \frac{NS^{2}}{2}((3J_{1}-6J_{2}+3J_{3})\notag\\
          &\qquad\qquad \pm \frac{1}{2}(-J_{4}+3J_{5}-6J_{6})),\notag\\
E_{z}^{I/II} -E_{0}&= \frac{NS^{2}}{2}((-J_{1}+2J_{2}+3J_{3})\notag\\
             &\qquad\qquad\pm \frac{1}{2}(-J_{4}-J_{5}+2J_{6})),\notag\\
E_{s/ferri}^{II}- E_{0}&=  \frac{NS^{2}}{4}((J_{1}-3J_{3})\notag\\
             &\qquad\qquad -(J_{4}-J_{5}-2J_{6})),\notag\\
E_\mathrm{z/ferri}^{I} -E_{0} &= \frac{NS^{2}}{4}((-J_{1}+3J_{3}),\notag\\
         &\qquad\qquad -(J_{4}+J_{5}-2J_{6})),\notag\\
E_{AFM-a}-E_{0} &= \frac{NS^{2}}{2}((-3J_{1}-6J_{2}-3J_{3})\notag\\
             &\qquad\qquad+\frac{1}{2}(J_{4}+9J_{5}+2J_{6})),\notag\\
\label{eq2}            
\end{align}

Here, I and II refer to Conf. I and Conf. II, respectively. The abbreviations z, s, ferri, and AFM-a stand for zigzag, stripy, ferrimagnetic (three atoms aligned ferromagnetically per unit cell), and a phase with two FM layers coupled antiferromagnetically (see Fig.~\ref{00fig0}(a)). Additionally, we have included various in-plane configurations that exhibit stabilities close to the zigzag configuration, such as zigzag-ferrimagnetic (z/ferri) and stripy-ferrimagnetic (s/ferri). SM Section~\ref{mixedconf} provides further details about these configurations. Moreover, $N$ is the total number of magnetic atoms per cell, $S$ is the spin (S $=$ 1), and (J$_{1}$, J$_{2}$, J$_{3}$) and (J$_{4}$, J$_{5}$, J$_{6}$) correspond to the intralayer and interlayer magnetic exchange interactions, respectively (see, Fig.~\ref{fig-J}). The total energies for these configurations calculated with the relaxed structures of the zigzag-zigzag configuration are given in Table \ref{tab:energy-BS1P}.\\

The M$_{P}^{S_1}$ stacking has two 1$^{st}$ and two 2$^{nd}$ NN, along with four 3$^{rd}$ NN interlayer exchange interactions (see Fig.~\ref{fig-J}).
For the zigzag configuration, the magnetic interaction between the first NN determines the AFM/FM magnetic coupling between layers (see J$_{4}$, depicted in bi-lines in Fig.~\ref{fig-J}). In contrast, the second NN does not contribute due to spin frustration (see J$_{5}$ depicted with solid line in Fig.~\ref{fig-J}). Moreover, the third NN contributes to both AFM and FM interactions. Therefore, for the zigzag configuration, both J$_{4}$ and J$_{6}$ contribute to Eq.~(\ref{ising1}) (see Eqs.~(\ref{eq1}), E$_{z}^{I/II}$).
However, the situation changes for other magnetic in-plane spin arrangements. For the N\'eel-N\'eel configuration, all interlayer NN are frustrated, resulting in no net interlayer exchange (see Eqs.~(\ref{eq1}), E$_{N\acute{e}el}$). In the case of stripy-ferri, the interaction with the first NN is frustrated (J$_{4}$), and the interaction between the second and third NN dominates the magnetic exchange(see Eqs.~(\ref{eq1}), E$_{s/ferri}$). Finally, the ferri configuration depends on the linear combination of J$_{4}$ and J$_{5}$ (see Eqs.~(\ref{eq1}), E$_{ferri}$).\\

On the other hand, in the M$_{P}^{TM}$ stacking, the A and B atoms in the honeycomb lattice of the metal atoms are no longer equivalent. For instance, in the zigzag-zigzag configuration (as shown in Fig.~\ref{fig-J}), the A atoms have one 1$^{st}$ NN, three 2$^{nd}$ and six 3$^{rd}$ NN (depicted with double red lines, a solid blue line and a dashed black line, respectively). In contrast, the B atoms have six 2$^{nd}$ NN, and the 3$^{rd}$ NN is beyond 8.7 \AA; therefore, it is not considered. Unlike the M$_{P}^{S_{1}}$  stacking, the M$_{P}^{TM}$  configuration does not exhibit frustration, regardless of the magnetic configuration. Therefore, the interlayer interactions depend on the linear combinations of J$_{4}$, J$_{5}$, and J$_{6}$ (see Eq.\ref{eq2}). \\

The Ising model, which defines exchange parameters based on various magnetic configurations and distances between metal atoms while considering interactions with different neighbors, may overlook  information related to interlayer interactions, where long-range distances have been shown to play a significant role \cite{jang2019microscopic,si2021revealing,sivadas2018stacking}.
In bilayer systems, neighboring atoms may have similar distances but follow different hopping paths, resulting in distinct couplings for the same separation. To account for these factors and provide a more comprehensive understanding of the relationship between Wannier orbitals and exchange parameters, we compute the exchange interactions by solving the Heisenberg Hamiltonian using the magnetic force theorem as implemented in the TB2J code \cite{tb2j}. 

\begin{figure}[]{\includegraphics[width=0.26\textwidth]{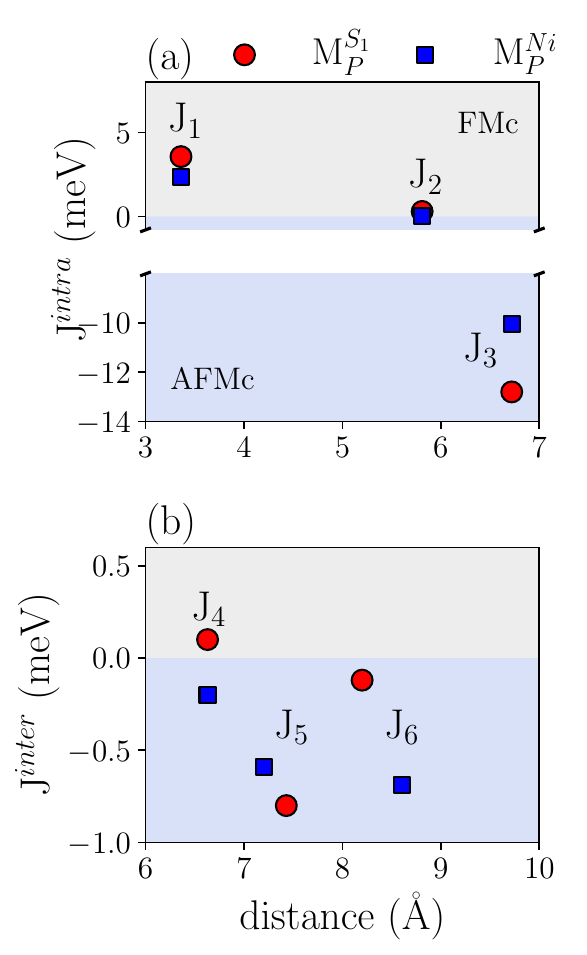}}{\includegraphics[width=0.23\textwidth]{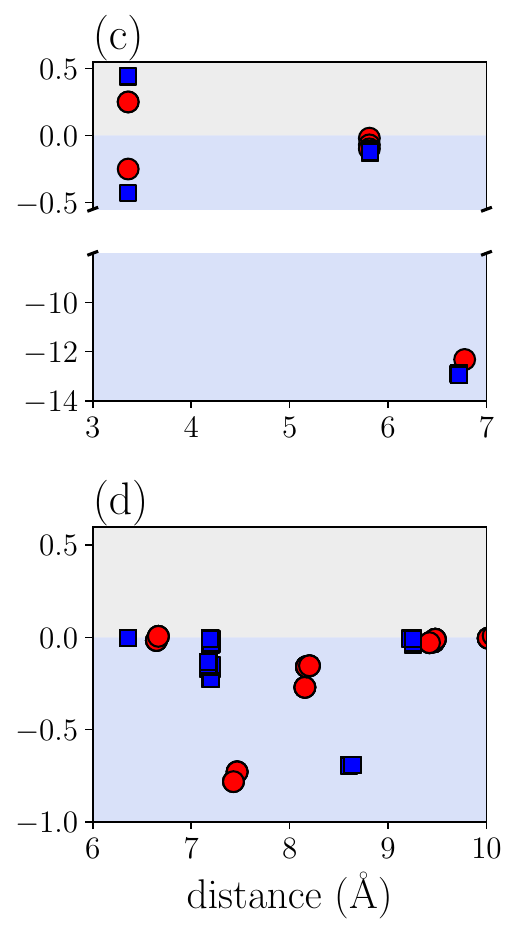}}
\caption{Magnetic exchange constants derived from the Ising Hamiltonian (Figs. (a-b)) and using the Magnetic Force Theorem as implemented in TB2J (Figs. (c-d)). Both models indicate that the interlayer magnetic exchange interaction is primarily contributed by the 2$^{nd}$ and 3$^{rd}$ nearest neighbors}.
\label{J-results}
\end{figure}

\begin{figure}[ht]
\centering
\includegraphics[width=0.4\textwidth]{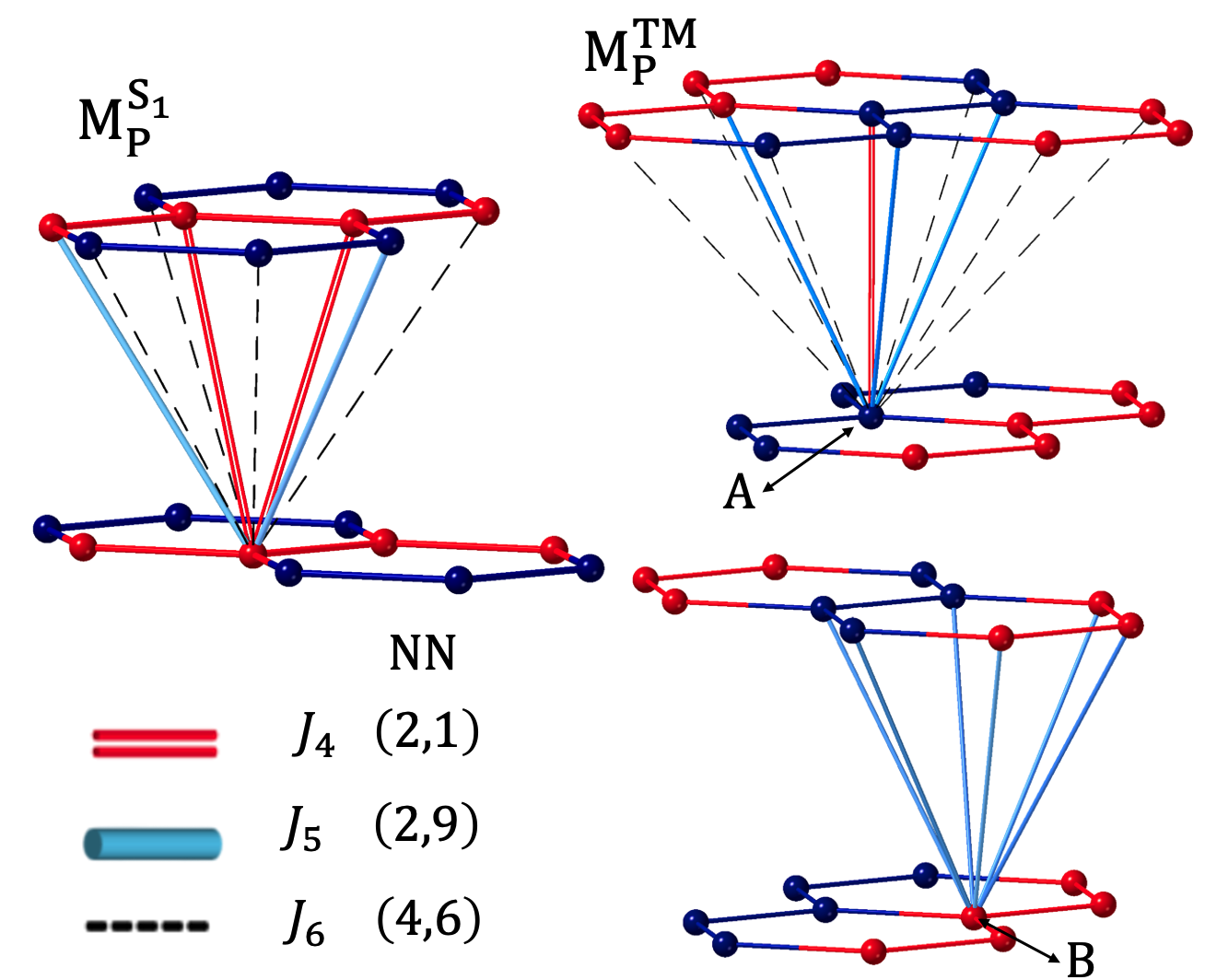}
\caption{Figure shows the M$_{P}^{S_{1}}$ and M$_{P}^{TM}$ stacking, respectively. It displays the bonds for first (red bi-lines), second (solid blue line), and third interalyer nearest neighbors (green dotted lines). In the case of M$_{P}^{TM}$, the interaction between the layers differs for the A and B atoms of the honeycomb lattice. Parenthesis numbers indicates the number of nearest neighbors (NN) for each stacking (M$_{P}^{S_{1}}$, M$_{P}^{TM}$).}
\label{fig-J}
\end{figure}

Figure~\ref{J-results} (a-b) show the resulting intralayer and interlayer exchange terms solving Eqs.~(\ref{eq1}-\ref{eq2}) and the derived with Heisenberg Hamiltonian. We find good agreement between both methods, with the exception of J$_{1}$
(which we will refer to later). In both stackings, the predominant intralayer exchange interaction J$^{intra}$ comes from the 3$^{rd}$ NNs at a distance of $6.75$ Å. Our findings for J$_{2}$ and J$_{3}$ align with those reported in previous studies utilizing a Ising model \cite{olsen2021magnetic} and real-space tight-binding Hamiltonian \cite{autieri2022limited}. However, our calculated J$_{1}$ $\sim$ 0.5 meV, from the Heisenberg Hamiltonian, is lower than previously reported values \cite{autieri2022limited}. This discrepancy likely arises from our inclusion of all intralayer J contributions without limiting them to the first three NN. \\

In NiPS$_{3}$, the Ni ion with a Ni$^{2+}$ valence state, possesses eight valence electrons in the $d$-orbital shell. The t$_{2g}$ orbitals are fully occupied in this electronic configuration, contributing predominantly to the lower energy states. In contrast, the e$_{g}$ orbitals (d$_{x^2-y^2}$ and d$_{z^2}$) contribute mainly to the upper part of the conduction band, as illustrated in Fig.~\ref{fig:dos}(b). Therefore, the exchange terms will be determined mainly by these orbitals.

For the intralayer magnetic exchange in both NiPS$_{3}$ stackings, the orbital-resolved interaction matrix shows that for the 1$^{st}$ NN, there is a weak
direct magnetic interaction (J $\sim$ 0.5 meV) originating from the d$_{x^2-y^2}$ and $d_{z^2}$ orbitals. In contrast, for the 2$^{nd}$ NNs, there is no direct interaction, and SE mechanisms are disfavored due to the lack of alignment of S($p$) orbitals, resulting in nearly zero exchange interactions.
However, for the 3$^{rd}$ nearest neighbors (NNs), the alignment of Ni(d$_{x^2-y^2}$)- S($p$) $\cdots$ S($p$)-Ni(d$_{x^2-y^2}$) orbitals favors the SE path, inducing a strong magnetic coupling of $J_3 \approx -13$ meV per atom. In this case, for the M$_{P}^{S_{1}}$ system, the orbital-resolved interaction matrix is given by

\begin{equation}
J_{3} \sim \begin{bmatrix}
 \begin{array}{c|ccc}
& J_{d_{xy}} & J_{d_{x^{2}-y^{2}}}  & J_{ d_{z^{2}}}  \\
 \hline
 J_{d_{xy}} &   -0.002     & -0.092       & 0.006  \\
 J_{d_{x^{2}-y^{2}}} &   -0.184     &  -11.848 & -0.004 \\
 J_{d_{z^{2}}} &  0.004 & 0.000   &  -0.184 \\
\end{array}
\label{J3}
\end{bmatrix},
\end{equation}
similar results are obtained for the M$_{P}^{Ni}$ stacking. The Wannier orbitals for the intralayer interaction are shown
in the SM, ~\ref{intralayer}.

\begin{figure*}[h]
{\includegraphics[width=1.0\textwidth]{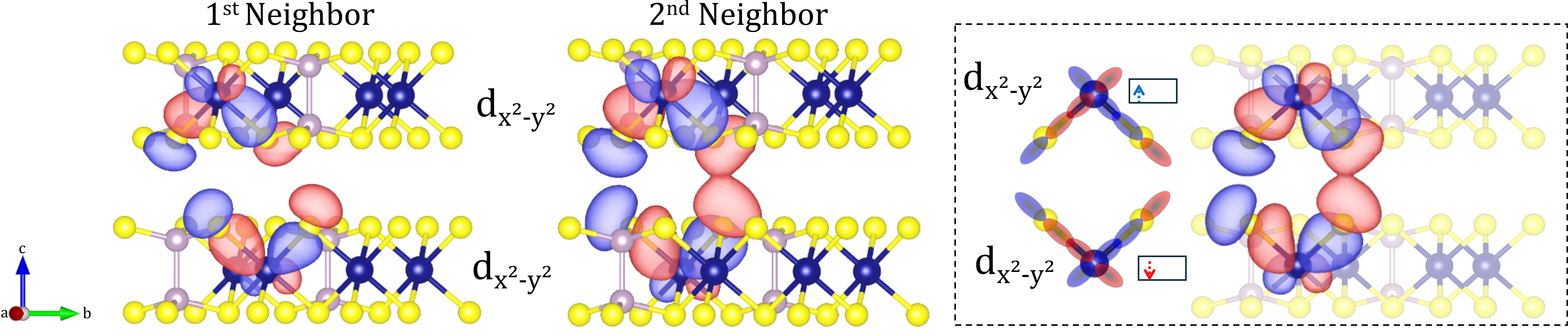}}
\caption{Localized Wannier orbitals for NiPS$_{3}$, M$_{P}^{S_{1}}$ stacking, for 1$^{st}$ and 2$^{nd}$ interlayer neighbors. For the 1$^{st}$ interlayer neighbors, there is no alignment between the p orbitals of the sulfur atoms in the upper and down layers, which happens with the 2$^{nd}$ interlayer neighbors. The right diagram shows a schematic representation of the interaction between 2$^{nd}$ interlayer neighbors. (Iso-surpericie with a value equal to 2.6).}
\label{Wannier-orbital}
\end{figure*}

In the case of the interlayer exchange terms, denoted as \(J^{\text{inter}}\), different trends can be observed, where the interaction can favor either a AFM or weak FM coupling depending on the distance for Ising results (Fig.~\ref{J-results}(b)). The M$_{P}^{S_1}$ stacking displays a weakly ferromagnetic coupling for the 1$^{st}$ NN located at 6.75 Å (J$_{4}$), with dominant magnetic interlayer interactions occurring at distances of 7.5 Å and 8.1 Å, corresponding to interaction strengths of approximately J$_{5}$ $=$ -0.8 meV (2$^{nd}$ NN) and J$_{6}$ $=$ -0.2 meV (3$^{rd}$ NN), respectively. However, a different trend is observed in M$_{P}^{Ni}$, where the strongest coupling occurs at around 8.6 Å (J$_{6}$ = -0.7 meV, 3$^{rd}$ NN). A similar trend is evident with the tight-binding model, although it favored AFM coupling independent of the nearest neighbor. Here, we see that the main contribution to \( J \) comes from the 2$^{nd}$ NN for M$_{P}^{S_1}$ and the 3$^{rd}$ NN for M$_{P}^{Ni}$. Nevertheless, the interactions with the 1$^{st}$ NN are strongly suppressed in both stackings (Fig.~\ref{J-results}(d)).

The strengthening of the magnetic exchange interaction between neighboring layers arises from the overlap of their orbitals through the super-superexchange interaction of Ni(d$_{z^{2}}$)-S(p) $\cdots$ S(p)-Ni(d$_{z^{2}}$) and Ni(d$_{x^{2}-y^{2}}$)-S(p) $\cdots$ S(p)-Ni(d$_{x^{2}-y^{2}}$) orbitals, resulting in an AFM nature, independent of the Ni coupling (AFM or FM). In the M$_{P}^{S_1}$ stacking configuration, the primary orbitals involved in the SE mechanism are the Ni(d$_{x^2-y^2}$) orbitals, which exhibit strong hybridization with the S(p) orbitals (see SM, Fig.~\ref{intralayer}). In the case of the 1$^{st}$ NN, the magnetic exchange interaction (J$_{4}$) is notably weakened due to the misalignment of the sulfur atoms' p orbitals in the upper and down layers. Conversely, for the 2$^{nd}$ NN, the alignment of these p orbitals between layers strengthens the superexchange (SE) mechanism, resulting in a more favorable magnetic exchange (J$_{5}$ $\sim$ -0.8 meV) (see, Fig.~\ref{Wannier-orbital}). In the M$_{P}^{Ni}$ stacking, we find that the magnetic interaction between the 1$^{st}$ and some of the 2$^{nd}$ NNs is strongly suppressed due to the lack of alignment of close S-S pairs. Moreover, J$_{6}$ is enhanced because there are three possible interactions involving the d$_{x^2-y^2}$ and $d_{z^2}$ orbitals at a distance of approximately 8.7 Å. 

The observed trend for the interlayer exchange interaction in NiPS$_{3}$ aligns with behaviors observed in other van der Waals magnets. Previous studies have shown that the primary contribution to interlayer exchange can originate from atom pairs located at distances longer than the first nearest neighbors. For instance, in compounds such as CrX$_{3}$ (X = Cl \cite{ebrahimian2023control}, Br \cite{si2021revealing} and I \cite{jang2019microscopic}), the strongest interlayer exchange can arise from either first or second neighbors, depending on the stacking configuration which in turn depends on SE mechanisms. 
 
In the case of FePS$_{3}$, we have detected some inconsistencies when comparing our Ising results with the tight-binding model based on WF analysis. By solving Eqs.~(\ref{eq1}), we found that all interlayer interactions favor AFM coupling, with J$_{5}$ $\sim$ J$_{3}$ (see further details in SM Table \ref{table-Jfe}). However, we also found that the strength of the interlayer exchange constants strongly depends on the magnetic configurations used to calculate them. In our tight-binding model analysis for bulk-type stacking, we obtained the same sign for all magnetic couplings (intralayer and interlayer), which contradicts the zigzag magnetic order. Due to the electronic structure of FePS$_{3}$, which includes singly-occupied d$_{xz}$/d$_{yz}$ and e$_{g}$ orbitals (see Fig.~\ref{fig:dos}), we anticipate a strong competition between ferromagnetic and antiferromagnetic interlayer interactions. This competition involves a combination of several orbital channels, which could result in weaker interlayer interactions. This scenario could explain the weak exchange energy observed in this system compared to NiPS$_{3}$ (see Fig.~\ref{fig2}(c-d)).\\

In summary, our study provides valuable insights into the interlayer magnetic interactions and stacking preferences in NiPS$_{3}$ and FePS$_{3}$ bilayers. NiPS$_{3}$ displays a weak AFM interaction between the d$_{x^{2}-y^{2}}$ and d$_{z^{2}}$ orbitals of the 1$^{st}$ NNs. The main interaction comes from the 2$^{nd}$ or 3$^{rd}$ NNs depending on the stacking, since the alignment of the sulfur orbitals of the two layers provides an SE interaction path, analog to the process that happens on the 3$^{rd}$ NN intralayer interaction \cite{autieri2022limited}.

In case of  FePS$_{3}$, our results suggest that antiferromagnetic coupling between layers is more energetically favorable for the M$_{P}^{S_1}$ (bulk-type stacking), in contrast, M$_{P}^{Fe}$ stacking favors ferromagnetic coupling. Although our findings align with experimental reports of the magnetic ground state in bulk-type stacking, the lowest energy stacking corresponds to the M$_{P}^{Fe}$ configuration. We observed a strong interplay between the stacking and magnetic configurations in both NiPS$_{3}$ and FePS$_{3}$ systems, which also depend on intralayer magnetic configurations.
 
Our studies indicate the potential tunability of magnetic coupling through stacking shifts in NiPS$_{3}$, making it relevant for exploring TMPS$_{3}$ moir\'{e} systems. Unlike Fe, Ni has strong magnetic exchange coupling for high-symmetry stacking. Therefore, neutron spectroscopy could detect changes in their magnetic configuration with stacking. The ability to control magnetic interactions in differently stacked regions opens up intriguing possibilities, particularly in emergent magnetic textures, spin polarization, or even the formation of skyrmions. Exploring TMPS$_{3}$ moiré structures could be an avenue to manipulate magnetic order and optical properties in 2D antiferromagnets, and could potentially lead to the exploration of altermagnetic phases as recently suggested \cite{liu2024twisted}. There is still much to investigate in TMPS$_{3}$ systems, and our study took the first step by conducting a theoretical analysis of the magnetic interactions in stacked antiferromagnetic bilayer systems.

\section{Acknowledgments}

We thank Efrat Lifshitz, Magdalena Birowska and Jhon Gonzalez for fruitful discussions. The authors gratefully acknowledge the computing time made available to them on the high-performance computer at the NHR Center of TU Dresden and on the high-performance computers Noctua 2 at the NHR Center PC2. These are funded by the German Federal Ministry of Education and Research and the state governments participating on the basis of the resolutions of the GWK for the national high-performance computing at universities (www.nhr-verein.de/unsere-partner). A.L. thanks to Chilean FONDECYT Postdoctoral Grant No. 3220505. We would like to thank the German Science Foundation (DFG) for supporting this work with the DIP grant No.1223-21. BC acknowledges financial support by the DFG via the SFB 1415, Project ID No. 417590517.

\newpage
\bibliographystyle{unsrtnat}
\bibliography{bib}

\pagebreak

\onecolumn
\begin{center}

\textbf{\large Supplemental Material:\\ Interlayer Magnetic Coupling in FePS$_{3}$ and NiPS$_{3}$ Stacked Bilayer\\}

\author{Andrea Le\'on,$^{1,4}$ Beatriz Costa,$^{2,3}$ Thomas Heine,$^{2,3,4,5}$ Thomas Brumme$^4$}

\date{%
$^1$Departamento de F\'{i}sica, Facultad de Ciencias, Universidad de Chile, Casilla 653, Santiago, Chile. \\%
$^2$Helmholtz-Zentrum Dresden-Rossendorf, Bautzner Landstr. 400, 01328 Dresden, Germany.\\
$^3$Center for Advanced Systems Understanding, CASUS, Untermarkt 20, 02826 Görlitz, Germany.\\
$^4$Chair of Theoretical Chemistry, Technische Universit\"at Dresden, Bergstrasse 66, 01069 Dresden, Germany.\\
$^5$Yonsei University and ibs-cnm, Seodaemun-gu, Seoul 120-749, Republic of Korea.\\
\today
}

\end{center}

\setcounter{figure}{0} 
\setcounter{section}{0} 
\setcounter{equation}{0}
\setcounter{table}{0}
\setcounter{page}{1}
\renewcommand{\thepage}{S\arabic{page}} 
\renewcommand{\thesection}{S\Roman{section}}   
\renewcommand{\thetable}{S\arabic{table}}  
\renewcommand{\thefigure}{S\arabic{figure}} 
\renewcommand{\theequation}{S\arabic{equation}} 

\section{Wannier functions}
\label{Wannier}

We employed Wannier localized functions for the tight-binding model, constructing the Wannier Hamiltonian with the d orbitals of the metal atoms and the p orbitals of the sulfur atoms. The localization of these functions was verified by examining the ratio between the real and imaginary components of the orbitals, ensuring it was close to zero. Additionally, we confirmed the alignment between the DFT and Wannier band structures and evaluated the orbital spreads, which were less than 1 Å for the d orbitals and less than 2 Å for the p orbitals. During the Wannierization process, 10,000 steps were conducted to ensure proper localization, although maximally localized Wannier functions were not utilized. The magnetic interactions were calculated using the TB2J code, which employs the Green's function method with local rigid spin rotation treated as a perturbation to derive the exchange parameters. For orbital analysis, the system was rotated so that the octahedra formed by the metal and sulfur atoms were aligned with the Cartesian axes, meaning the layers were not perpendicular to the z-axis.

\section{(Fe,Ni)PS3 Bulk}\label{SM-bulk}

\begin{table*}[h]
\centering
    \begin{tabular}{c| c c c c c}
\hline
 & $a$ (\AA) & $b$  (\AA) & $c$  (\AA) & d$_{L}$  (\AA) &  \\
\hline
\hline
NiPS$_{3}$-Computed   & 5.81 & 10.09  & 6.62  & 6.32  & \\
NiPS$_{3}$-Experiment & 5.80 \cite{liang2017general}& 10.03 \cite{liang2017general}  & 6.60 \cite{liang2017general} & 6.34 \cite{pazek2024charge}    &
\\
\hline
FePS$_{3}$-Computed   & 5.98 & 10.30  & 6.73 & 6.45 & \\
FePS$_{3}$-Experiment &  5.95 \cite{budniak2022spectroscopy}&  10.30 \cite{budniak2022spectroscopy}  & 6.72 \cite{budniak2022spectroscopy} & $\sim$ 6.40 \cite{yan2022layer}    &
\\
\hline
\end{tabular}
    \caption{Lattice parameter and layer distance (d$_{L}$).}
\label{table-Jfe}
\end{table*}

\begin{figure*}[ht]
\centering
{\includegraphics[width=0.3\textwidth]{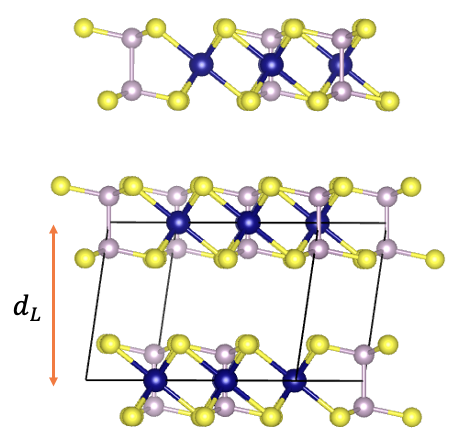}}
\caption{FePS$_{3}$ and NiPS$_{3}$ Bulk.}
\end{figure*}

\section{Bilayer stability as Hubbard-U parameter}\label{Uvalues}

Table~\ref{tab:Energy-U} shows the energy difference between the lowest energy stackings (M$_{P}^{S_{1}}$ and M$_{P}^{TM}$ for FePS$_{3}$ and NiPS$_{3}$ system, respectively) for several Hubbard-U values U = 4, 5 and 6 eV. Hence, the lowest energy stacking is independent of the U value in the explored range.

\begin{table*}[h]
\begin{center}
\resizebox{18cm}{!}{
\begin{tabular}{ c | c | c | c | c | c | c }
\hline
\multicolumn{1}{c|}{Stackings} & \multicolumn{3}{c|}{FePS$_{3}$} & \multicolumn{3}{c}{NiPS$_{3}$} \\ 
\hline
\multicolumn{1}{c|}{} & \multicolumn{1}{c|}{$\Delta$E (meV), U = 4 eV} & \multicolumn{1}{c|}{$\Delta$E (meV), U = 5 eV} & $\Delta$E (meV), U = 6 eV & \multicolumn{1}{c|}{$\Delta$E (meV), U = 4 eV} & \multicolumn{1}{c|}{$\Delta$E (meV), U = 5 eV} & $\Delta$E (meV), U = 6 eV \\ 
\hline
M$_{P}^{S_{1}}$  & 10.18 & 6.98 & 7.95 & 0 & 0 & 0  \\
M$_{P}^{TM}$     & 0   & 0 & 0 & 14.56 & 12.63 & 10.95 \\
\hline
\end{tabular}}
\caption{Energy difference between the M$_{P}^{TM}$ and M$_{P}^{S_{1}}$ stackings, for Fe and Ni, considering three Hubbard-U values (U = 4, 5, and 6 eV).} 
\label{tab:Energy-U}
\end{center}
\end{table*}

\begin{figure*}[ht]
\centering
{\includegraphics[width=0.9\textwidth]{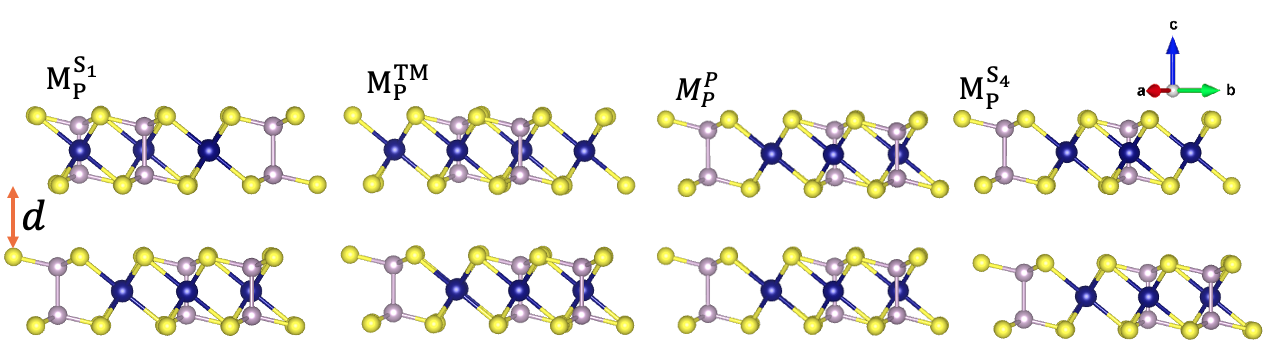}}
\caption{Front view of selected high symmetry stacking. ``$d$" is the distance between S$^{top}$-S$^{down}$ layers.}
\label{front-bilayer}
\end{figure*}

\section{Magnetic configuration}\label{magconf}

\begin{figure*}[ht]
\centering
{\includegraphics[width=0.8\textwidth]{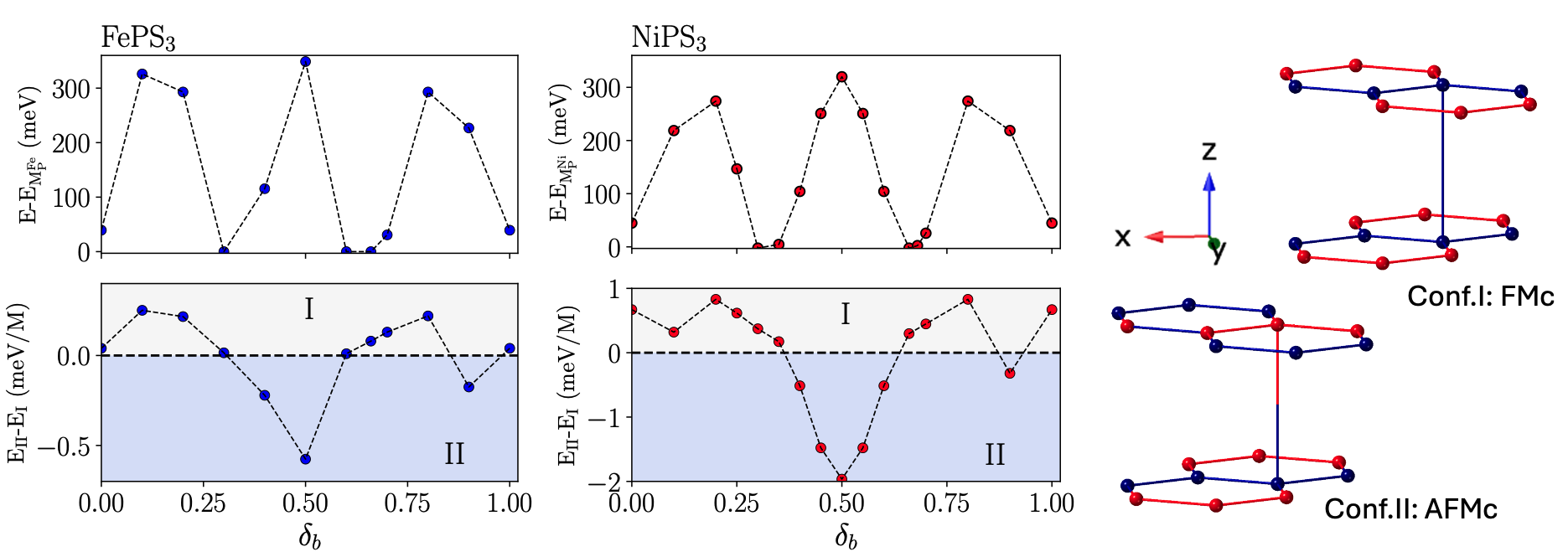}}
\caption{Upper panel: Stacking energy as a function of the shift along the $b$-direction. Lower panel displays the energy difference between Conf.II (II) and Conf.I (I), which represent two possible magnetic couplings between the layers (see Fig.~(c)), defined as antiferromagnetic and ferromagnetic interactions among first neighbors in M$^{M}_{P}$ stacking ($\delta$ = 2/3).}
\label{fig:diff_b}
\end{figure*}

Figure~\ref{fig:diff_b}(a)-(b) upper panel illustrates the energy stacking due to a shift by $\delta$ along the [010] direction for FePS$_{3}$ and NiPS$_{3}$, respectively. The lower panel displays the energy differences between Conf.II and Conf.I. (Conf.I and Conf.II correspond to FM and and AFM respectively, as shown in Fig.~\ref{fig:diff_b}(c)).

\subsection{Interplay between in plane and interlayer configurations}\label{mixedconf}

\begin{figure*}[ht]
\centering
{\includegraphics[width=0.5\textwidth]{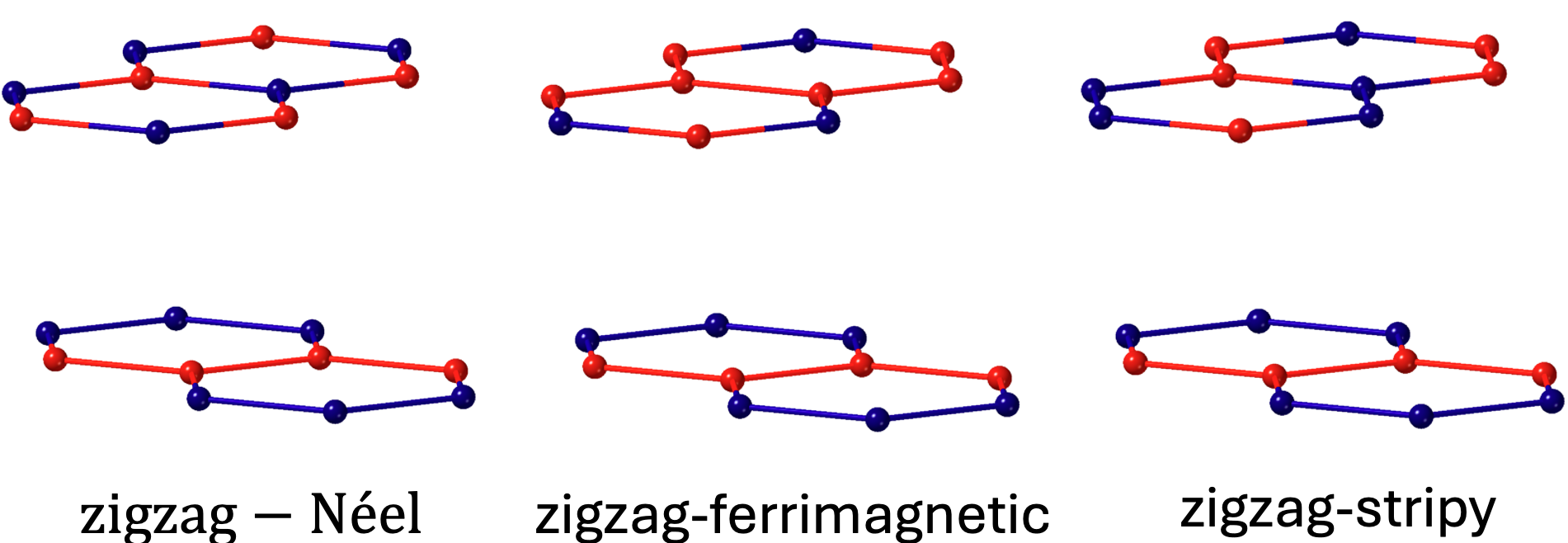}}{\includegraphics[width=0.5\textwidth]{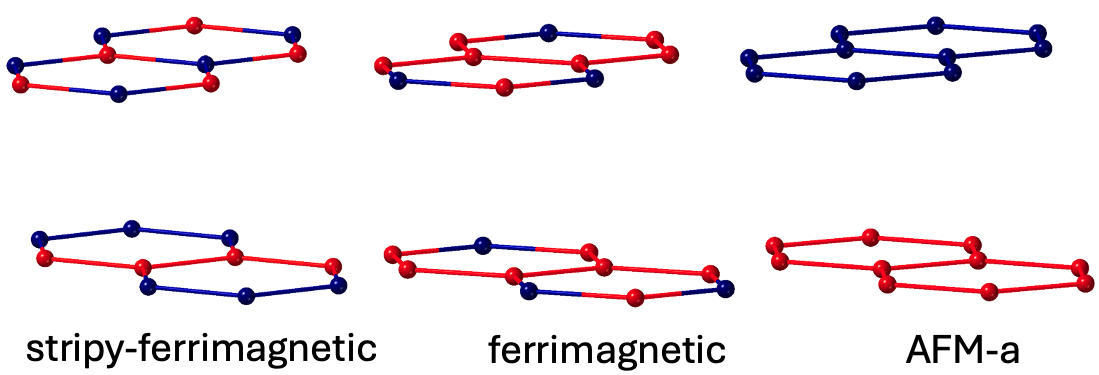}}
\caption{Schematic representation of TMPS$_{3}$ with mixed magnetic configurations between layers and an AFM-a phase (two FM layers coupled antiferromagnetically). The ferrimagnetic configuration consists of three atoms aligned ferromagnetically per unit cell.}
\label{fig:mix}
\end{figure*}

\begin{table}[ht]
\centering
    \begin{tabular}{ c | c | c }
         &   BL-FePS$_{3}$ (ML)  &   BL-NiPS$_{3}$ (ML)  \\        
\hline
 Configuration        & $\Delta$E (meV) & $\Delta$E  (meV)  \\
\hline
zigzag      &    0  (0)      &   0  (0) \\ 
 Neél       &    3.98  (3.85)   &   6.07    (3.59) \\  
 ztripy     &    13.29 (12.098) &   41.87   (30.20)  \\
 zig-Neél   &    0.90  &    3.14       \\
 zig-stripy &   8.07     &   20.86      \\
 zig-ferri   &    3.59   &  6.04        \\
 ferri      &     -- (5.68)      &   19.93   (15.13) \\
 stripy-ferri   &  --        &  31.27     \\
 FM         &    20.82   (25.03)  &    32.73   (24.80)\\
 \hline
\end{tabular}
    \caption{Energy difference among different magnetic configurations concerning zigzag state ($\Delta$E = E$_{conf.}$-E$_{zigzag}$) in meV/M for the bilayer (BL) system with M$_{P}^{S_1}$ stacking. The label ``-'' means that the system is unstable, and the value in the parenthesis corresponds to the system in the monolayer case.}
\label{table-energy}
\end{table}

We explored different magnetic configurations such as N\'eel, stripy, and FM, all of which have already been investigated in monolayer structures. Additionally, we included combinations of these magnetic configurations based on zigzag-configurations, such as zigzag-N\'eel, zigzag-stripy, and zigzag-ferrimagnetic (ferri) (see Fig.~\ref{fig:mix}). Table~\ref{table-energy} shows the energy differences among the different configurations for Fe and Ni systems with M$_{P}^{S_1}$ (bulk-type stacking). For both systems, zigzag has the lowest configuration energy, as expected. However, we found that the zigzag-N\'eel configuration follows as the next metastable state, followed by configurations with higher energy such as N\'eel, zig-stripy, and zig-ferrimagnetic configurations. Additionally, we observed that the stripy configuration is the most unfavorable state. However, when combined with zigzagigzag, this configuration gains stability. Other mixed magnetic configurations based on the N\'eel-configuration are metastable states with higher energy. Interestingly, we observed that the combining zigzag with ferrimagnetic coupling is much more favorable than other antiferromagnetic configurations, such as stripy and even ferromagnetic coupling. This observation holds beyond the bilayer, as demonstrated in the monolayer case.

\subsection{NiPS$_{3}$: stacking magnetic configurations.}
\label{energy}
Table~\ref{tab:energy-BS1P} shows the total energy for the lowest magnetic configuration for both stackings. In the case of M$_{P}^{S_1}$ stacking, the configurations used include zigzag-I/II, N\'eel, zigzag-ferri, ferri-I, and stripy-ferri-I/II. In the case of the M$_{P}^{Ni}$ stacking, the configurations used are zigzag-I/II, N\'eel-I/II, zigzag-ferri-I, and stripy-ferri-II  and AFM-a (notation I and II are the abbreviations of Conf.I and Conf. II, respectively). Fig. \ref{fig:mix} shows a schematic representation of each configuration.

\begin{table}[h!]
\centering
\begin{tabular}{c| c| c| c}
\hline
M$_{P}^{S_{1}}$-Configurations & Energy (eV) &  M$_{P}^{Ni}$-Configurations & Energy (eV) \\
\hline
 zigzag-I      & -112.706    & zigzag-I         &  -110.321 \\
 zigzag-II      &  -112.703   & zigzag-II       &  -110.319  \\
 N\'eel-I       &  -112.657   &   N\'eel-I      &   -110.278  \\
zigzag-ferri-I  &  -112.624  &  N\'eel-II       &   -110.288 \\  
ferri-I         & -112.538   &  zigzag-ferri-I  &   -110.256   \\
stripy-ferri-II &  -112.457  &  stripy-ferri-II  &  -110.129  \\
stripy-ferri-I  &  -112.452   &   AFM-a            & -110.129  \\
\hline
\end{tabular}
\captionof{table}{NiPS$_{3}$: Total energy for M$_{P}^{S_{1}}$ and M$_{P}^{Ni}$  stacking.} 
\label{tab:energy-BS1P}
\end{table}

\subsection{FePS$_{3}$ stacking magnetic configurations:}\label{SM:Fe}

Table \ref{tab:energy-BMP}  shows the total energy for the lowest magnetic configuration for both stackings. For M$_{P}^{S_1}$,  zigzag-I/II, N\'eel, zigzag-ferri-I, ferri-I and stripy-ferri-I. In case of M$_{P}^{Ni}$, we find zigzag-I/II, N\'eel-I/II, zigzag-ferri-I/II, stripy-I/II.\\

Table \ref{tab:table-Jfe} shows the intralayer (J$_{1}$, J$_{2}$ and J$_{3}$) and interlayer (J$_{4}$, J$_{5}$ and J$_{6}$) magnetic exchange, for  M$_{P}^{S_{1}}$, by solving the eq.\ref{eq1} (shown in the main text) and for M$_{P}^{Fe}$ solving the set of equations given by \ref{eq-3}.

\begin{table}[h]
    \centering
\begin{tabular}{c| c| c| c}
\hline
M$_{P}^{S_{1}}$-Configurations & Energy &  M$_{P}^{Fe}$-Configurations & Energy \\
\hline
 zigzag-I    &  -137.803    & zigzag-II    &  -137.813 \\
 zigzag-II    &  -137.802    & zigzag-I    &  -137.812  \\
 N\'eel-I      &  -137.771      &   N\'eel-I  &   -137.782  \\
zigzag-Ferri-I  &  -137.762     &  N\'eel-II &   -137.781 \\  
ferri-I         & -137.703      & stripy-II   &   -137.641   \\
stripy-ferri-II &  -137.656  & stripy-II  &   -137.628  \\
stripy-ferri-I &  -137.622  & FM &   -137.745   \\
\hline
\end{tabular}
\captionof{table}{FePS$_{3}$: Total energy for M$_{P}^{S_{1}}$ and M$_{P}^{Ni}$ stacking.} 
\label{tab:energy-BMP}
\end{table}

Table \ref{tab:table-Jfe} shows the magnetic exchange for FePS$_{3}$ by solving Eqs.~(\ref{eq1}) and Eqs~.(\ref{eq-3}) using Table \ref{tab:energy-BMP} values. 

\begin{table*}[h]
\centering
    \begin{tabular}{c|c c c c c c c}
\hline
& Stacking & J$_{1}$ & J$_{2}$ & J$_{3}$ & J$_{4}$ & J$_{5}$ & J$_{6}$ \\
\hline
\hline
FePS$_{3}$  & M$_{P}^{S_1}$  & 1.82  & 0.6  & -1.96  & -0.06 & -1.10 & -0.046\\
            & M$_{P}^{TM}$   & 1.15 & 0.33  & -1.29 & -0.02 & -0.22 & -0.12\\
\hline
\end{tabular}
    \caption{Intralayer (J$_{1}$, J$_{2}$, J$_{3}$) and interlayer (J$_{4}$, J$_{5}$, J$_{6}$) exchange interaction in meV for the lowest energy stacking.}
\label{tab:table-Jfe}
\end{table*}

\begin{align}
E_\mathrm{N\acute{e}el}^{I/II} &= E_{0}-\frac{NS^{2}}{2} ((-3J_{1}+6J_{2}-3J_{3})\pm \frac{1}{2}(-J_{4}+3J_{5} -6J_{6})),\notag\\
E_{z}^{I/II} &= E_{0}- \frac{NS^{2}}{2}((J_{1}-2J_{2}-3J_{3}) \pm \frac{1}{2}(J_{4}+J_{5}-2J_{6})),\notag\\
E_{s}^{I/II} &= E_{0}- \frac{NS^{2}}{2}((-J_{1}-2J_{2}+3J_{3})\pm\frac{1}{2}(J_{4}-3J_{5}-2J_{6})),\notag\\
E_\mathrm{FM} &= E_{0}-\frac{NS^{2}}{2}((3J_{1}+6J_{2}+3J_{3}) +(J_{4}+9J_{5}+6J_{6})),
\label{eq-3}
\end{align}

\subsection{Wannier orbitals: NiPS$_{3}$ Intralayer}\label{intralayer}

\begin{figure*}[ht]
\centering
{\includegraphics[width=0.75\textwidth]{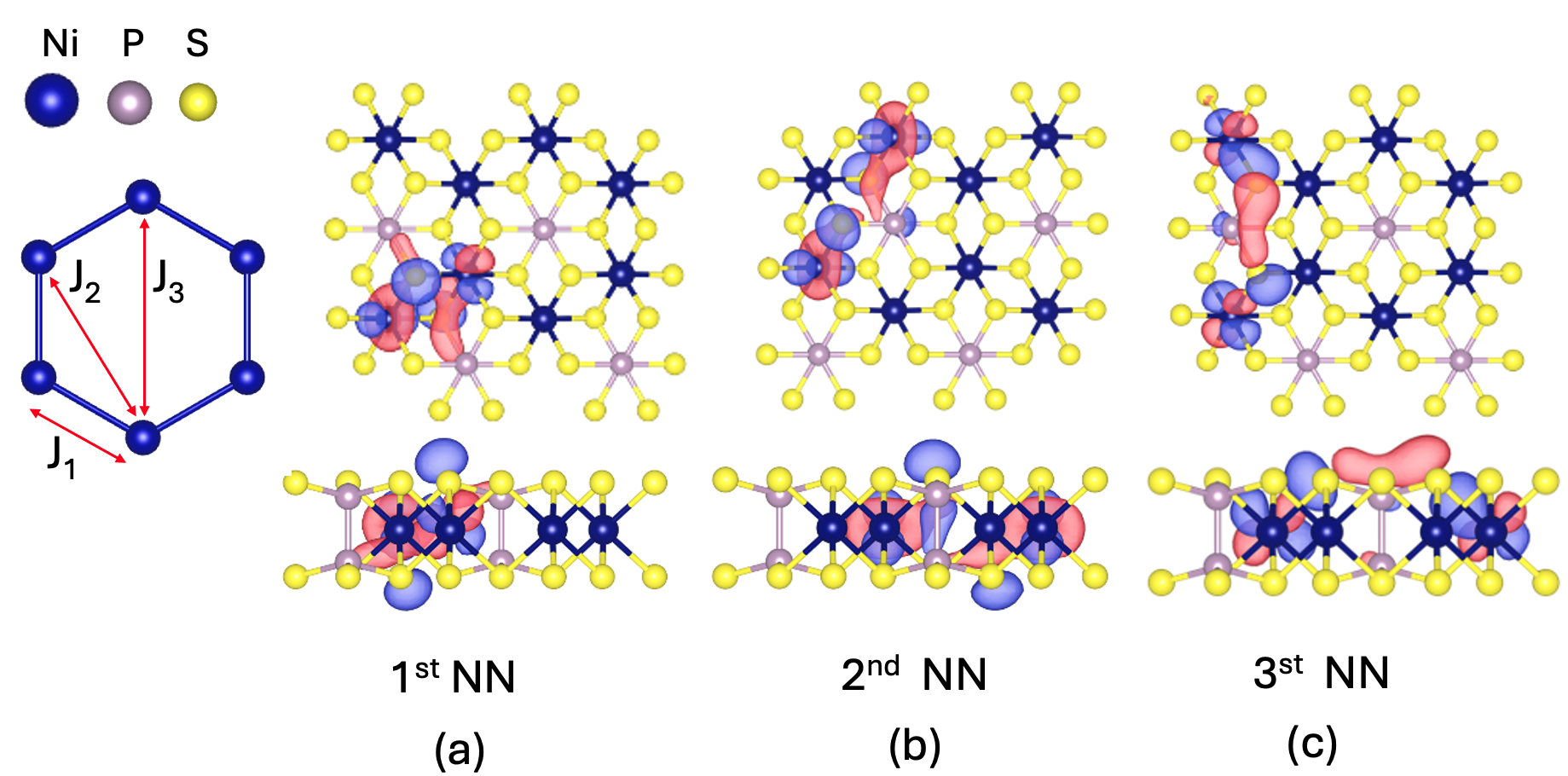}}
\caption{Localized Wannier orbitals for NiPS$_{3}$ intralayer nearest neighbor. (a), (b) and (c) 1$^{st}$, 2$^{nd}$ and 3$^{rd}$ neighbor, respectively.}
\label{intralayer0}
\end{figure*}

Figure \ref{intralayer0} shows the main intralayer Wannier orbitals involved for 1$^{st}$, 2$^{nd}$, and 3$^{rd}$ intralayer NN.

\end{document}